\documentclass[namedreferences]{solarphysics}

\usepackage[hyperref,optionalrh]{spr-sola-addons} 
\usepackage{graphicx}        
\usepackage{color}           
\usepackage[T1]{fontenc}

\newcommand{\eg}{{e.g.,~}}

\newcommand\arcsec{\mbox{$^{\prime\prime}$}}%

\chardef\us=`\_

\begin{document}

\begin{article}
\begin{opening}

\title{Multiwavelength observations of a breakout jet at an active region periphery}

\author[addressref={aff1,aff2},corref,email={virat.com@gmail.com}]{\inits{P.}\fnm{Pradeep}~\lnm{Kayshap}}
\author[addressref=aff3,email={judith.t.karpen@nasa.gov}]{\inits{J.T.}\fnm{Judith T.}~\lnm{Karpen}}
\author[addressref={aff3,aff4},email={pankaj.kumar@nasa.gov.in}]{\inits{P.}\fnm{Pankaj}~\lnm{Kumar}}

\address[id=aff1]{School of Advanced Sciences and Languages, VIT Bhopal University, Kothrikalan, Sehore, Madhya Pradesh - 466114}
\address[id=aff2]{Inter University Centre for Astronomy $\&$ Astrophysics, Post Bag - 4, Ganeshkhind, Pune 411007, India}
\address[id=aff3]{Heliophysics Science Division, NASA Goddard Space Flight Center, Greenbelt, MD 20771, USA}
\address[id=aff4]{Department of Physics, American University, Washington, DC 20016, USA}

\runningauthor{Kayshap et al.}
\runningtitle{Breakout jet at an active region periphery}

\begin{abstract}
We analysed Interface-Region Imaging Spectrograph (IRIS) and the Solar Dynamics Observatory/Atmospheric Imaging Assembly (SDO/AIA) observations of a small coronal jet that occurred at the solar west limb on 2014 August 29. The jet source region, a small bright point, was located at an active-region periphery and contains a fan-spine topology with a mini-filament. Our analysis has identified key features and timings that motivate the following interpretation of this event.  As the stressed core flux rises, a current sheet forms beneath it; the ensuing reconnection forms a flux rope above a flare arcade.  When the rising filament-carrying flux rope reaches the stressed null, it triggers a jet via explosive interchange (breakout) reconnection. During the flux-rope interaction with the external magnetic field, we observed brightening above the filament and within the dome, along with a growing flare arcade. EUV images reveal quasi-periodic ejections throughout the jet duration with a dominant period of 4 minutes, similar to coronal jetlets and larger jets. We conclude that these observations are consistent with the magnetic breakout model for coronal jets.  
\end{abstract}
\end{opening}

\section{Introduction}
     \label{S-Introduction} 
The collimated flow of plasma in the form of solar jets is an integral part of the solar atmosphere, with important consequences for coronal dynamics and energetics. Solar jets occur on various spatio-temporal scales throughout the solar atmosphere  (e.g., \citealt[and references therein]{Shibata1992, Shimojo1996, Innes1997,  Sterling2000, Savcheva2007, Shibata2007, Paraschiv2015, Jelinek2015, Raouafi2016, Kayshap2018, Kayshap2021}).
The collective behaviour of the magnetic field and plasma form the various types of solar jets (e.g., spicules, macrospicules, coronal hole jets, active-region jets, and network jets). Observations from the soft X-ray telescope (SXT) onboard Yohkoh showed a wealth of coronal jets, Since then the technology has improved continuously, and we are in an era of high-resolution solar observations from radio wavelengths to hard X-rays that have shown the wealth of solar jets on all scales, and advanced our understanding of the jet triggering mechanism and context.



A variety of reconnection-driven models for jets has been proposed: flux emergence, mini-filament eruption, flux cancellation \citep[\eg ][]{Young2014, Panesar2016, McG2019}, internal reconnection within a kinked flux tube \citep{Kayshap2013a, Kayshap2013b}, resistive-kink--driven interchange reconnection, and the magnetic breakout model. Although these models differ in the way sufficient free magnetic energy is built up to enable eruption, all incorporate some form of interchange reconnection between the closed field of the jet source and the external field. In addition, all of these models rely on the embedded-bipole (also known as fan-spine) topology as the fundamental structure in which the energy buildup and release occur. In resistive-kink models, twisting motions in the photosphere build magnetic stress, which finally leads to ideal instability; reconnection between the twisted flux and external open flux releases the jet into the solar atmosphere \citep[\eg ][]{pariat2009, pariat2010, Fang2014, pariat2015, pariat2016, Karpen2017}).  
The well-studied and simulated magnetic breakout model explains the formation of jets in fan-spine topologies through the same reconnection-driven mechanism that has been proposed for larger solar eruptions (CMEs/eruptive flares) \citep{Antiochos1998, Antiochos1999, Lynch2008, Karpen2012, Wyper2017}.  The mini-filament eruption paradigm \citep[\eg ][]{Shen2012, Sterling2015, Shen2017}) is essentially the same as breakout, as both involve formation and eruption of a filament channel inside a fan-spine configuration. Evidence for the key signatures of the breakout scenario in jets --- mini-flare arcades, mini-filament eruptions, fan footpoint brightenings, and plasmoids in the flare and breakout current sheets --- has been revealed by analyses of SDO and IRIS data \citep{Kumar2018, Kumar2019a, Kumar2019b} and simulations \citep{Wyper2016b, Wyper2017, Wyper2018}.
As we demonstrate in this paper, the event studied here also is consistent with the magnetic breakout model. 

Comparing the results of two- and three-dimensional (3D) magnetohydrodynamic (MHD) numerical simulations with the ground truth of observations offers the most rigorous tests of competing theories. In addition to identifying the underlying magnetic-energy storage and release mechanisms, establishing the evolving plasma and magnetic properties of jets through model---data comparisons is important for determining their global impact. For example, one key characteristic of most coronal jets that models must reproduce is their rotating motions \citep[\eg ][]{patsou2008, Zhang2014, Cheung2015b, Moore2015, Kayshap2018, Kumar2018}, which provides important clues about the underlying energy buildup and release processes. Here we study the evolution, kinematics, and triggering mechanism of a jet that occurred at the limb on 2014 August 29. Section~\ref{sect:obs_data} discusses the jet observations. In Section~\ref{sect:results}, we present the results of our data analysis. Our discussion and conclusions are stated in Section~\ref{sect:discussion}.

\section{Observations and Data Analysis}
\label{sect:obs_data}
We have utilized high-resolution imaging observations from the Interface-Region Imaging Spectrograph (IRIS) and the Atmospheric Imaging Assembly (AIA) onboard the Solar Dynamics Observatory (SDO) to study a jet occurring around 06:00:00~UT on 2014 August 29 in NOAA Active Region 12146 on the west limb of the Sun. AIA obtains full disk images of the Sun in the following chromospheric- and coronal-temperature channels: 94~{\AA}, 131~{\AA}, 171~{\AA}, 193~{\AA}, 211~{\AA}, 304~{\AA}, 1600~{\AA}, and AIA~1700~{\AA} \citep{Lemen2012}. AIA has a spatial resolution of 0.6 arcsec per pixel with a cadence of 12 s. The 3D noise-gating technique \citep{deforest2017} was used to clean the AIA images.

 This event was observed by IRIS in 8-step raster mode. IRIS provides high-resolution slit-jaw images (SJI) in ~2796~{\AA}, ~1330~{\AA}, and ~1400~{\AA} filters \citep{DePon2014}. This event was observed by IRIS/SJI only in the 1330~{\AA} channel, which mainly captures transition-region-temperature emission. The spatial resolution of IRIS/SJI is 0.33 arcsec with a cadence of 10 s in this observation. We have utilized the level 2 data files from both instruments that are ready for scientific purposes. We have investigated the evolution of this jet using the IRIS~1330~{\AA}, AIA~1600~{\AA}, AIA~304~{\AA}, and AIA~94~{\AA} images. The jet was located near the edge of the IRIS field-of-view (FOV), so IRIS only captured the major dynamics at the base of the jet. The entire jet was observed by AIA, however. 
 
\section{Analysis} \label{sect:results}

\begin{figure*}
    \mbox{
    \includegraphics[trim = 3.5cm 0.1cm 2.5cm 0.1cm,scale=0.29]{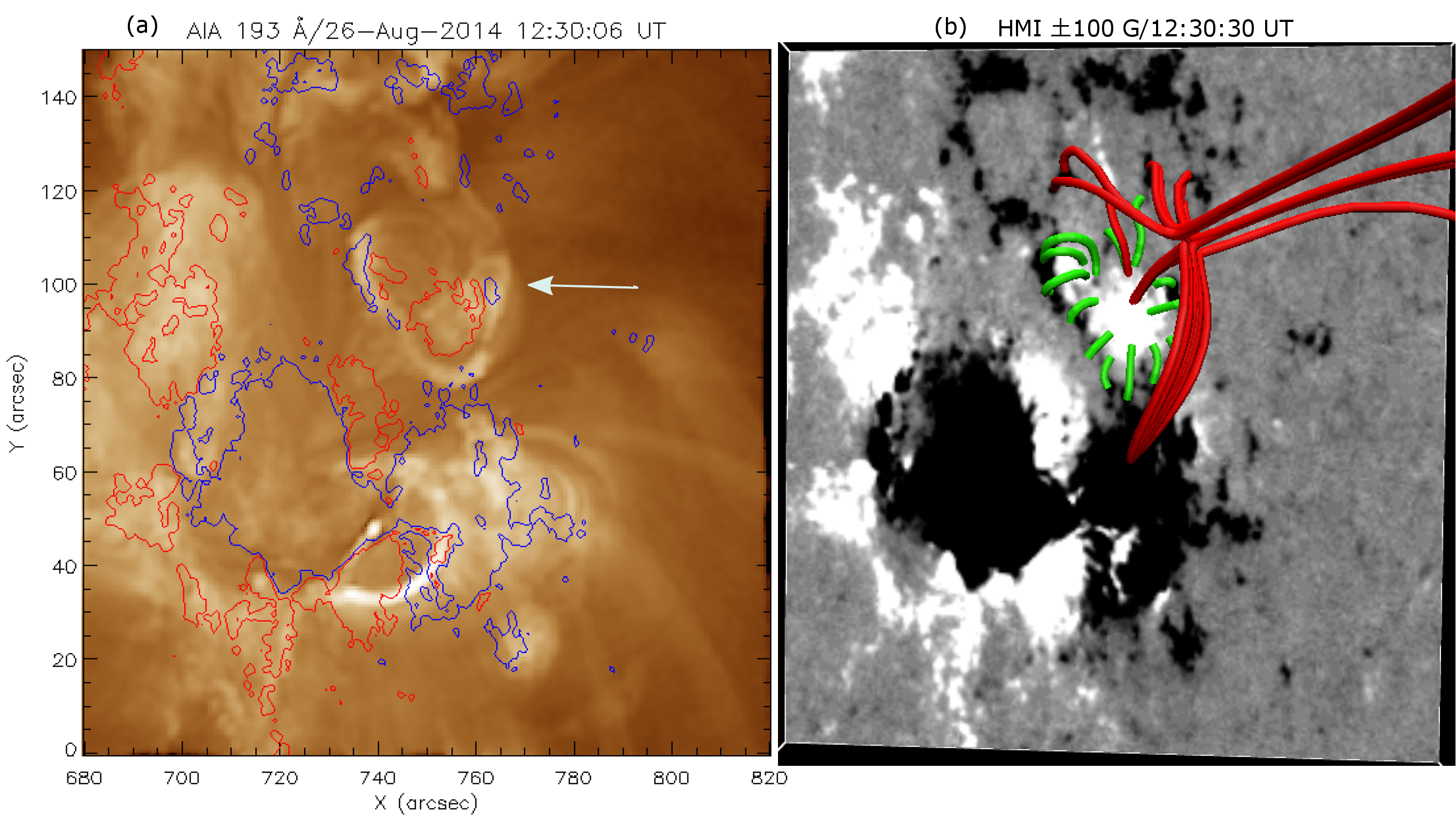}
    }
    \caption{(a) AIA 193 {\AA} image overlaid by HMI magnetogram contours ($\pm$100 G, red=positive, blue=negative) on 26 August 2014 (3 days before the jet). The arrow points to the coronal bright point that is the jet source. The panel (b) shows the cotemporal HMI magnetogram where white indicates positive polarity and black indicates negative polarity. The overplotted green and red lines show the magnetic field configuration extracted with the help of potential field extrapolation code.}
    \label{fig:ref_fig}
\end{figure*}
\subsection{Magnetic Configuration of the Source Region} \label{sect:mag_field}
The jet occurred at the solar west limb on 2014 August 29 at the periphery of NOAA Active Region (AR) 12146, when the AR was slightly on the other side of the west limb.  Therefore, no photospheric magnetograms of the source region were available on the same day.  
Hence, we utilized SDO's Helioseismic and Magnetic Imager \citep[HMI;][]{scherrer2012} line-of-sight (LOS) magnetogram and EUV images of AR12146 at about 12:30 UT on 2014 August 26 (3 days prior to the jet event) to estimate the magnetic field configuration and associated coronal plasma structures in the vicinity of the jet region.\\

\begin{figure*}
    \mbox{
    \includegraphics[trim = 1.2cm 0.1cm 2.5cm 0.1cm,scale=0.80]{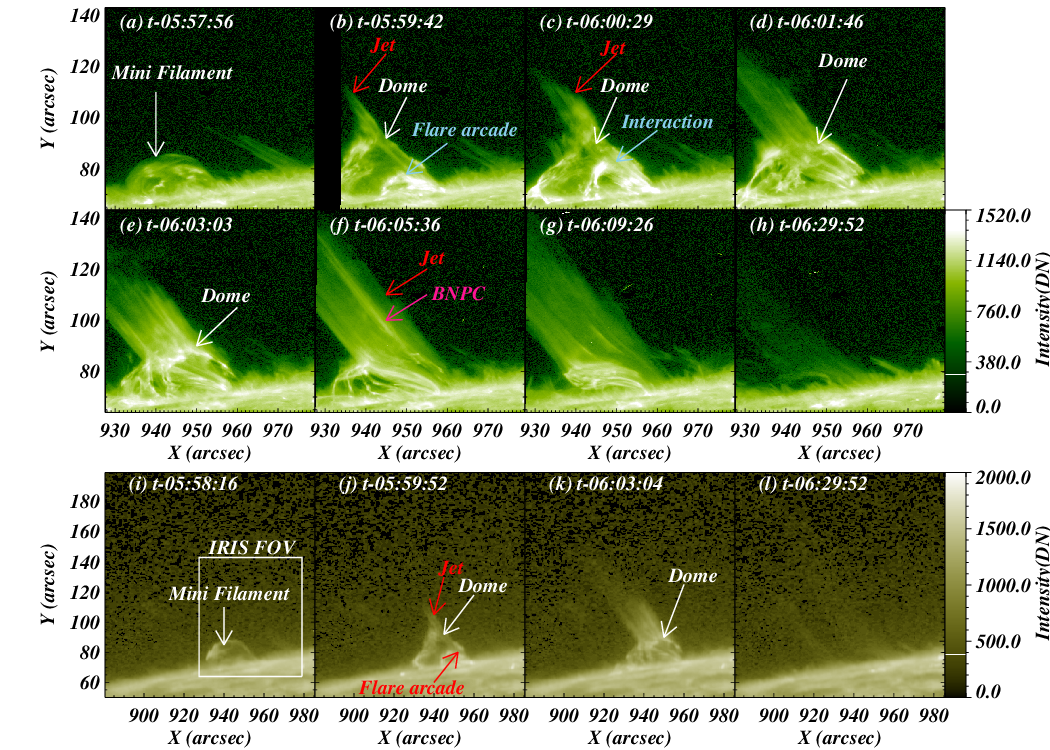}
    }
    \caption{Jet evolution in cool temperature channels: IRIS/SJI 1330~{\AA} (panels (a-h) and AIA 1600~{\AA} (panels (i-l). The jet was triggered around 05:58~UT (panel b), and around 06:30~UT, the jet vanished completely (panels h and l). During the jet's lifetime, we have observed various important features, namely, mini-filament (panel a), flare arcades (panel b), dome (panels b, c, d, e, j, and k), and BNPC (panels f)). Please note that the field-of-view (FOV) of AIA~1600 images (i.e., panels i to l) is laregr than that of IRIS~1330~{\AA}, and the white box in panel (i) shows the IRIS/SJI FOV.}.
    \label{fig:jet_evol_cool}
\end{figure*}

Figure~\ref{fig:ref_fig}(a) shows the AIA~193~{\AA} image while Figure~\ref{fig:ref_fig}(b) shows the HMI LOS magnetogram of AR 12146. The red (positive polarity) and blue (negative polarity) contours on the coronal intensity map outline field strengths of $\pm$100 G, shown in panel (b) as white and black areas respectively. 
The jet source region, a coronal bright point at the edge of the AR,is indicated by white arrows in the AIA~193~{\AA} intensity map. The bright point (width$\approx$30$\arcsec$) has a dome-shaped structure (Figure~\ref{fig:ref_fig}(a)), which is clearly visible at the limb three days later (Figure~\ref{fig:jet_evol_cool}(a)). To determine the magnetic configuration of the source region, we utilized a potential-field extrapolation code \citep{nakagawa1972} from the GX simulator package of SSWIDL \citep{nita2015}.  The potential field extrapolation shows a classic fan-spine topology (Figure ~\ref{fig:ref_fig}(b)). Note that multiple jets were detected from the same active region during August 26-29, some of which originated in the same bright point. In addition, a similar jet associated with a filament eruption was observed about 40 minutes prior to the jet studied here. We focus on the current jet due to the limited availability of high-resolution IRIS observations.

\subsection{Evolution of the Jet} \label{sect:evol_jet}
In this section, we describe the evolution of the jet in IRIS/SJI 1330~{\AA} and AIA~1600~{\AA}, AIA~171~{\AA}, and AIA~94~{\AA} images, which cover photospheric to coronal temperatures.
Figure~\ref{fig:jet_evol_cool} shows the evolution of the jet in the cool IRIS/SJI 1330~{\AA} (panels (a-h) and AIA~1600~{\AA} filters (panels (i-l)) (the accompanying animation only shows the IRIS 1330~{\AA} images). Note that IRIS has a limited FOV which does not capture the whole extent of this jet event, while AIA provides full-disk images. Hence, for all AIA channels including 1600~{\AA}, we have used a larger FOV than for IRIS.
In Figure~\ref{fig:jet_evol_cool}(i), the white rectangular box outlines the IRIS FOV shown in panels (a-h). At time t = 05:57:56~UT, a mini-filament is clearly visible just above the west limb (white arrow in Figure~\ref{fig:jet_evol_cool}(a)). The mini-filament also appears in AIA~1600~{\AA} (panel (i)) and rises slowly (see the animation accompanying Figure 2).

Around t=05:59:42~UT, the dome is clearly visible in IRIS/SJI 1330~{\AA} (white arrow in panel (b)) and AIA~1600~{\AA} (white arrow in panel (j)) images along with the jet at the top of the dome (red arrow in panels (b) and (j)). 
A bright arcade (labeled ``Flare Arcade" in panel (b)) appears towards the right side of the dome.
 IRIS and AIA animations reveal rotation of the jet plasma, with the typical characteristics of a helical jet.

The bright arcade grows steadily thereafter, and meets with the nearest side of the dome around 06:00:29 (sky blue arrow in Figure~\ref{fig:jet_evol_cool}(c)), accompanied by further brightening of the dome and localized brightenings at the base. 

Meanwhile, the jet extends outward from the top of the dome and becomes wider (red arrow in panel (c)). The jet front has left the IRIS FOV by 06:01:46 (panel (d)). By t = 06:03:03~UT, the dome has expanded in width but is beginning to shrink in height, while the jet continues to broaden and lengthen (Figure~\ref{fig:jet_evol_cool}(e) and (k)). At t  = 06:05:36~UT, the jet (red arrow in Figure~\ref{fig:jet_evol_cool}(f)) has almost reached its maximum width. A very bright, thin, and long plasma column appeared within the jet at t = 06:05:36~UT, which is indicated by a pink arrow in Figure~\ref{fig:jet_evol_cool}(f). Hereafter, this feature will be referred to as the bright and narrow plasma column (BNPC). 
The late phase of this event is characterized by the near disappearance of the bright dome and arcade and the persistence of the jet (Figure~\ref{fig:jet_evol_cool}(g)). Finally, the jet disappears completely around 06:30~UT  (Figure~\ref{fig:jet_evol_cool}(h) and (l)).    

AIA~171~{\AA} observations reveal the behaviour of $\approx$1 MK plasma. 
\begin{figure*}
    \includegraphics[trim = 2.5cm 4.0cm 6.8cm 0.0cm,scale=1.0]{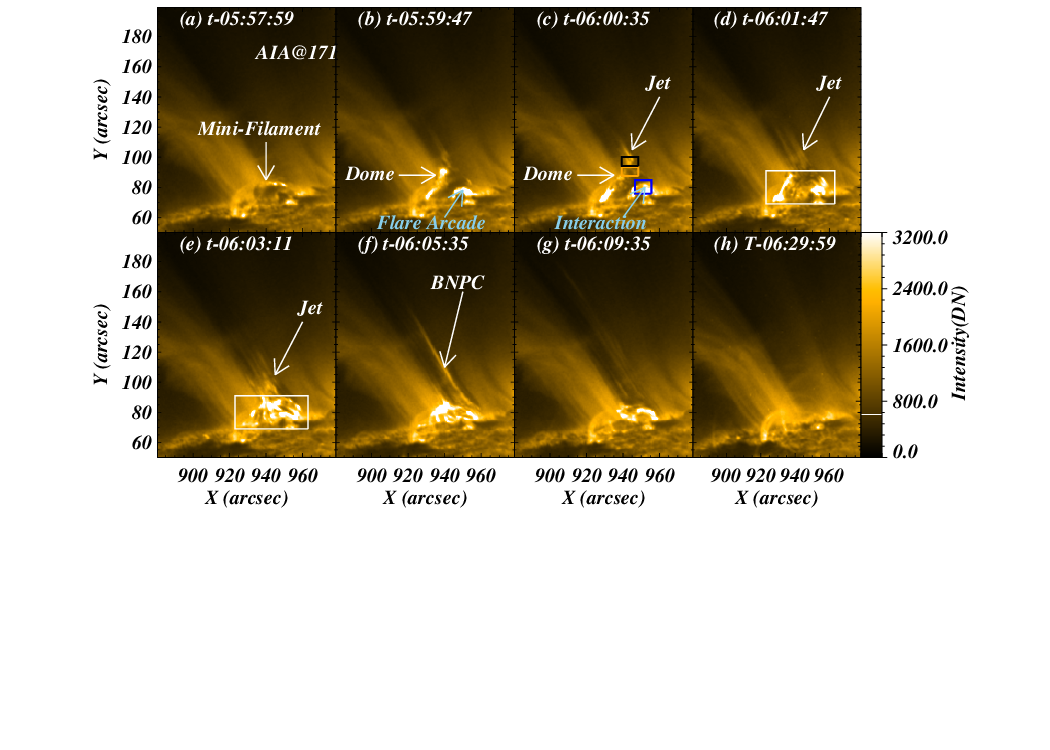}

    \caption{(a-h) The jet evolution in AIA 171~{\AA}. All important features as seen in Figure~\ref{fig:jet_evol_cool} (i.e., mini-filament, flare arcade, and BNPC) are also visible in AIA~171~{\AA} filter. Here, we have used white boxes in panels (d) and (e) to outline the dome. The three boxes in panel (c) outline areas in the jet spire (black box), the top of the dome (orange box), and a region on the right side of the dome (blue box). We utilized these regions to investigate the temporal variability (Figures~\ref{fig:qpp_304} and \ref{fig:qpp_171}.) A combined animation of Figures 3, 4, and 5 is available as Supplemental Material.} 
    \label{fig:jet_evol_171}
\end{figure*}
The plasma emission outlining the magnetic-field configuration is most clearly visible in this channel (Figure~\ref{fig:jet_evol_171}(a)). The morphology of the dome and jet 
suggests that the underlying field has a fan-spine magnetic topology, as confirmed by a potential field extrapolation from an HMI magnetogram 3 days earlier (Figure \ref{fig:ref_fig}). 
In general, the morphology and dynamics of the jet in this warm filter are similar to the cool-temperature filters (Figure~\ref{fig:jet_evol_cool} and accompanying animation): (1) existence of the dome containing a mini-filament about 40 min prior to the expulsion of the jet (Figure~\ref{fig:jet_evol_171}(a)), (2) interaction between the field supporting the mini-filament and the dome (Figure~\ref{fig:jet_evol_171}(b)), (3) formation of a bright arcade that brightens and expands with time (Figure~\ref{fig:jet_evol_171}(b) and (c)), (4) interaction between the bright arcade and the right side of the dome (Figure~\ref{fig:jet_evol_171}(c)), and (5) localized brightenings of the dome at the base of the jet (Figure~\ref{fig:jet_evol_171}(b)-(e)). 

The mini-filament appears as a dark absorption feature in AIA~171~{\AA}, which is indicated by the white arrow in Figure \ref{fig:jet_evol_171}(a). We see the signature of a very faint, small jet at the top of the dome from t = 05:59:47-06:03:11~UT (see white arrows in Figure \ref{fig:jet_evol_171}(b){--}(e)). Because this faint jet is initiated around the same time and from the same location as the jet seen in IRIS/SJI~1330~{\AA} and AIA~1600~{\AA} (Figure~\ref{fig:jet_evol_cool}), it is probably part of that dynamic feature. 
Several brightenings were distributed over the dome from the initiation of the faint jet up to its maximum phase (see panels (b){--} (e)). 
The mini-filament erupts completely around 06:01:47~UT (panel (d)), accompanied by strong brightenings all over the dome (panels (d) and (e)). 

The BNPC is marked by the white arrow in Figure~\ref{fig:jet_evol_171}(f), in the same location as in the IRIS/SJI~1330~{\AA} (Figure~\ref{fig:jet_evol_cool}(d)). The late decay phase of the jet event is shown in  Figure~\ref{fig:jet_evol_171}(g-h). Figure~\ref{fig:jet_evol_171}(c) shows three boxes: an orange box near the top of the dome, a blue box near the right base of the dome, and a black box within the jet spire. We utilized these three regions to investigate the spatial and temporal variability of emission during this event in the AIA~304~{\AA} and 171~{\AA} filter observations (Figures~\ref{fig:qpp_304} and \ref{fig:qpp_171}), as discussed in Section~\ref{section:jet_qpp}.\\

\begin{figure*}
    \includegraphics[trim = 2.5cm 4.0cm 6.8cm 0.0cm,scale=1.0]{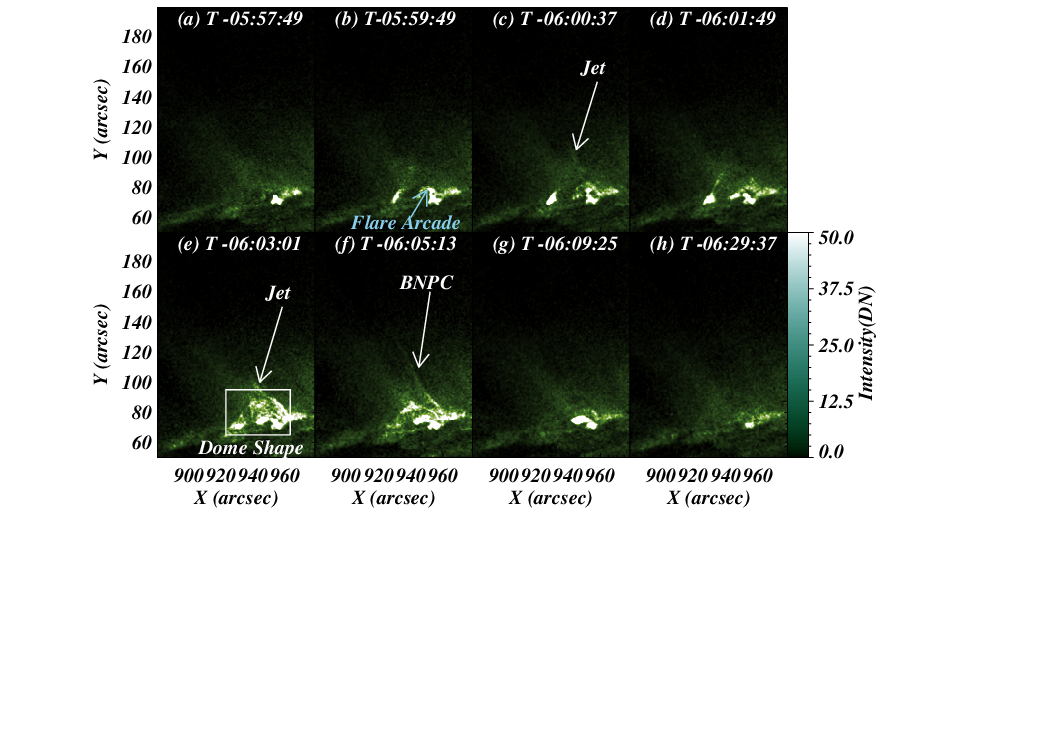}

    \caption{Successive AIA images of the Fe~{\sc xviii} emission (the 94 \AA\ channel minus the cool component), showing the jet evolution at hot temperatures $\geq$5 MK. Similar to Figure~\ref{fig:jet_evol_171}, the cyan arrow in panel (b), the white box in the panel (e), and white arrow in panel f shows/outlines the flare arcade, dome, and BNPC, respectively. A combined animation of Figures 3, 4, and 5 is available as Supplemental Material.}
    \label{fig:jet_evol_hot}
\end{figure*}
The AIA~94~{\AA} channel mainly captures emission from the hot plasma but may include emission from cool plasma as well \citep[\eg ][]{Lemen2012}. On the basis of extreme ultraviolet imaging and spectroscopic data, \citet{DelZanna2013} have proposed the following empirical formula that can remove the cool component from AIA~94~{\AA} observations by means of the AIA~171~{\AA} and AIA~211~{\AA} channels:

    I (Fe XVIII) = I(AIA~94) - I(AIA~211)/120 - I(AIA~171)/450 .\\    
    
With this equation, we have removed the cool component from the AIA~94{\AA} images and examined the evolution of the jet in the remaining hot emissions (Figure~\ref{fig:jet_evol_hot}). The Fe~{\sc xviii} emission (i.e., only hot emission from the 94~{\AA} filter; hereafter we will use the term Fe~{\sc xviii emission instead of 94~{\AA}) image} shows the flare arcade (panel (b)), fan dome (outlined by the white rectangular box in panel (e)), and very faint jet (panel (e); indicated by white arrow). The very high-temperature emission exhibits a faint signature of the jet between 06:01~UT and 06:05~UT (jet indicated by white arrows in panels (c) and (e)). Similar to cooler filters (e.g., IRIS/SJI~1330~{\AA} and AIA~171~{\AA}), the Fe~{\sc xviii} emission brightens progressively over the dome with time (Figure~\ref{fig:jet_evol_hot}(b-e) and accompanying animation). This progressively higher Fe~{\sc xviii} emission around the dome with time justifies the heating of the dome. Nearly the entire dome is bright around 06:03:01~UT, which is consistent with the 171~{\AA} image. The BNPC (indicated by the white arrow in panel (f)) appears within the jet in the Fe~{\sc xviii} emission images around 06:04:44~UT, consistent with all other analyzed IRIS/SJI and AIA/SDO filters.

\subsubsection{Event Summary}
Based on the observations described above, we have estimated the onset times of key features of this event, which are tabulated in Table~\ref{table:onset_time}. The dynamic evolution of the jet and its source in AIA~304~{\AA}, AIA~171~{\AA}, and Fe~{\sc xviii} emission is best seen in the animation accompanying Figure 3.
\begin{table}
\hspace{-0.5cm}
\begin{tabular}{ |p{1cm}|p{3cm}|p{3cm}|  }
 \hline
 \multicolumn{3}{|c|}{Onset time of key features} \\
 \hline
 Sr No.  & Event      & Onset Time\\
 \hline
 1.      & Interaction between mini-filament and dome & 05:57:55 \\
 \hline 
 2.      & Jet onset & 05:58:43\\
 \hline
 3.      & Appearance of mini-flare arcades in AIA~304~{\AA}  & 05:59:19 \\
 \hline
 4.      & Interaction between flare arcade and dome  & 06:00:07 \\
 \hline
 5.      & Onset of bright and narrow plasma column (BNPC) within the jet & 06:04:44 \\
 \hline
\end{tabular}
\caption{Onset time of key features.}
\label{table:onset_time}
\end{table}
The jet is triggered around t = 05:58:43~UT, after the interaction between the mini-filament and dome at 05:57:55~UT. Then, the flare arcades appear around 05:59:19~UT. The brightenings on and in the dome evolve, such that the whole dome is covered with brightenings around 06:03:01~UT. The BNPC forms on the right side of the jet at 06:04:44~UT.

\subsection{Time-distance plots}
\label{sect:td_plot}
We have performed a time-distance analysis of this jet in three AIA channels:
304~{\AA}, 171~{\AA}, and the Fe~{\sc xviii} emission extracted from the 94~{\AA} channel. 
\begin{figure*}
    \includegraphics[trim = 1.5cm 3.0cm 2.5cm 3.0cm,scale=0.8]{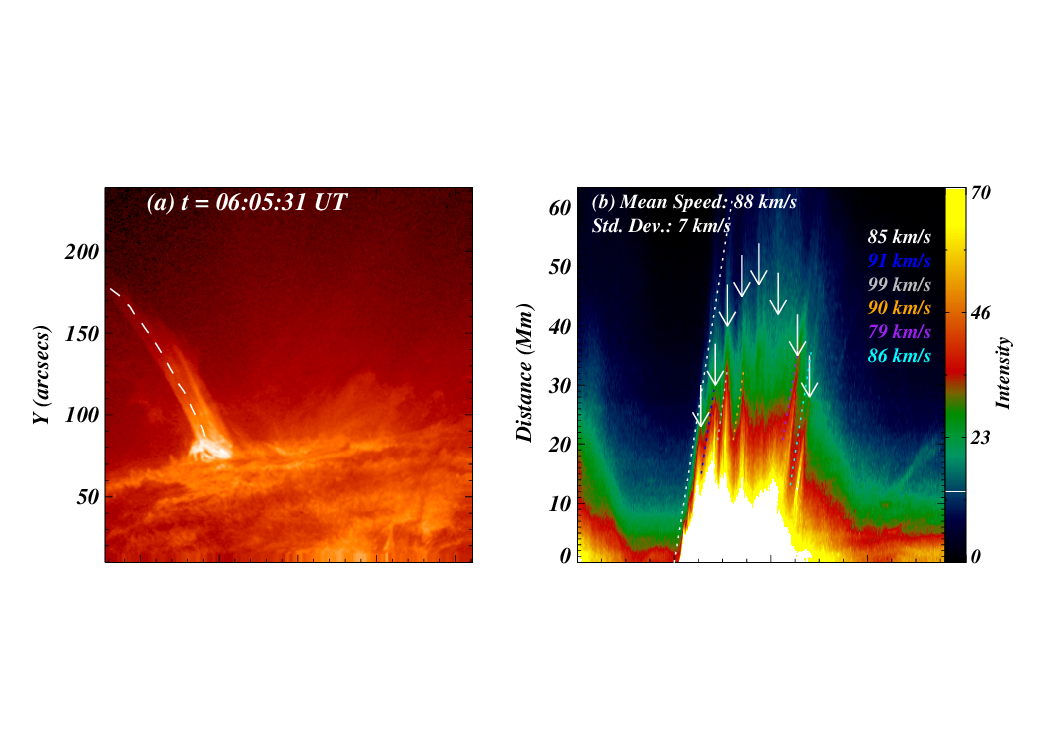}
    \includegraphics[trim = 1.5cm 3.0cm 2.5cm 3.0cm,scale=0.8]{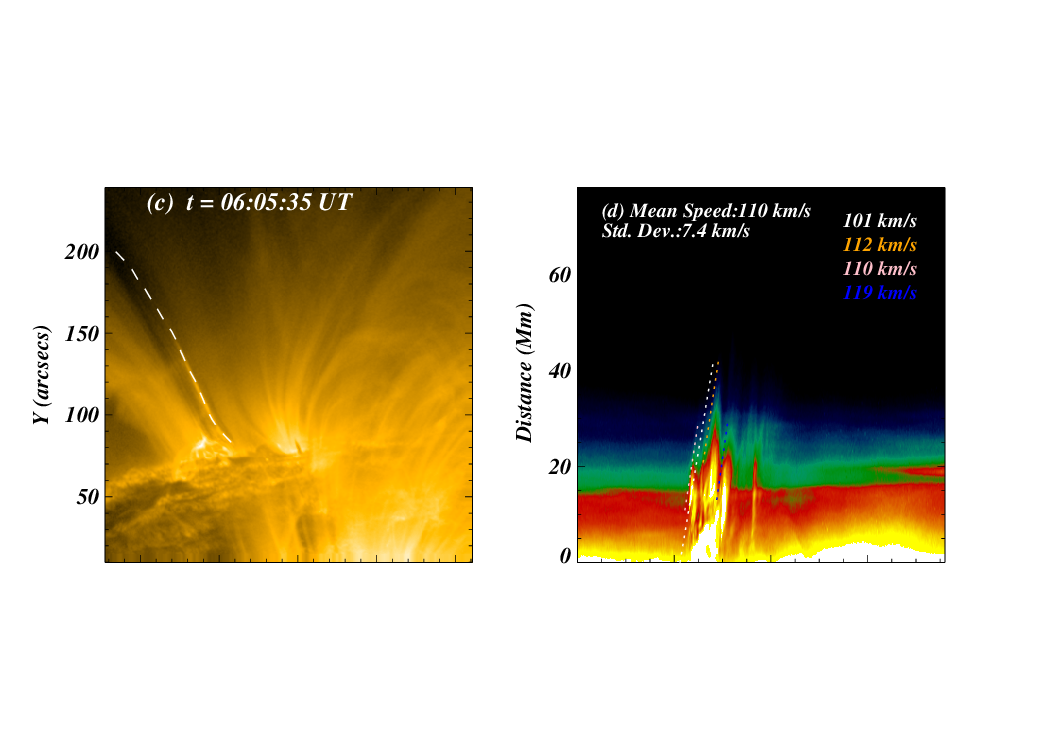}
    \includegraphics[trim = 1.5cm 2.0cm 2.5cm 3.0cm,scale=0.8]{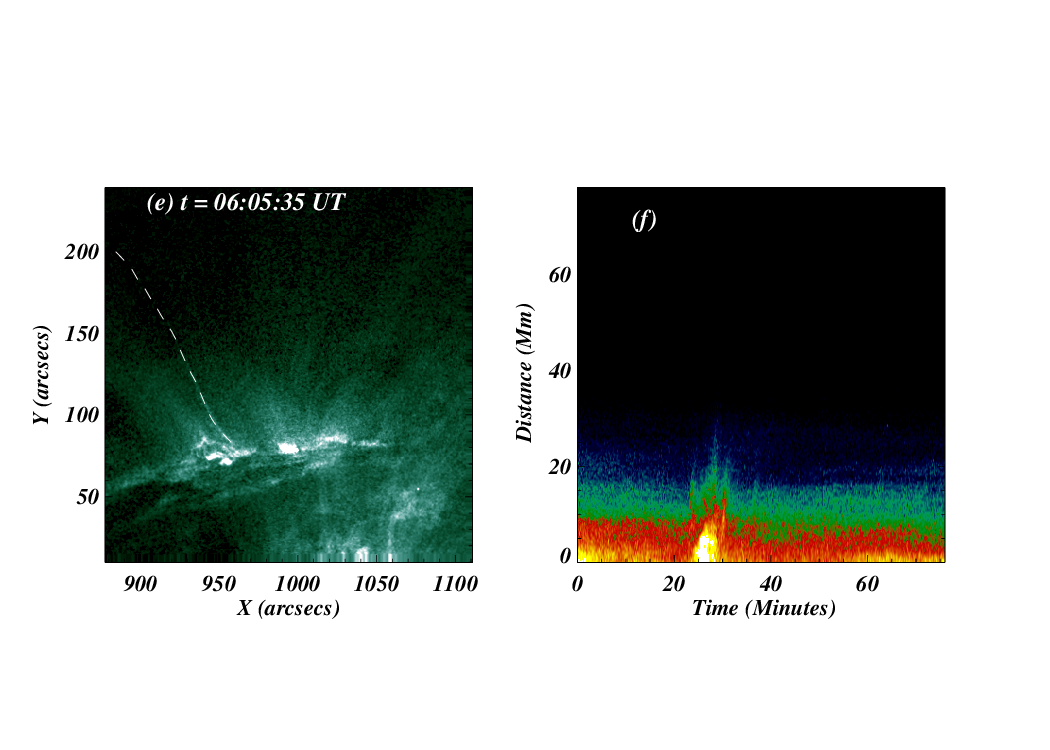}

   \caption{Reference images (left column) showing the slit location along the jets (white dashed lines) and corresponding time-distance plots (right column) in (a, b) AIA~304~{\AA}, (c,d) AIA~171~{\AA}, and (e,f) Fe~{\sc xviii} emission. We selected two different paths to produce time-distance plots: one path for AIA 304 ~{\AA} (panel (a)) that goes through the middle of the jet, and another for the warm/hot filters (panels (c) and (e)) that goes through the BNPC. The reference time t = 0 in all TD plots corresponds to 05:37:00~UT. The multiple paths (see various dotted lines) in the TD plots (b,d) are utilized to estimate the projected speeds of jet. The estimated speeds are written on the right side of TD plots. Finally, the average projected speed and corresponding standard deviation are estimated, and they are mentioned in the upper left corners of panels (b) and (d)}
   
 \label{fig:ht}
\end{figure*}  
The white dashed line drawn in Figure \ref{fig:ht}(a) marks the slit used to derive the time-distance (TD) plot of the jet from the AIA 304~{\AA} images. 
The animation accompanying Figure~\ref{fig:jet_evol_171} 
shows that the jet plasma ascended without any significant signatures of plasma downfall in the AIA channels. Because AR 12146 is located on the west limb of the Sun, the jet plasma could have fallen onto the far footpoints of the magnetic field lines along which the jet propagates, located behind the limb. 

The TD-plot corresponding to the white-dashed path on AIA~304~{\AA} filter is shown in panel (b). It should be noted that various jet-like features are evident during this jet event which are indicated by white arrows. We have drawn six different paths along the ascending slope of the jet in the TD plot, namely, see white-, blue-, gray-, orange-, purple-, and cyan-dashed paths on the TD plots (panel (b)). And, for each drawn path, we estimated the projected speeds, and the projected speeds are 85 km s$^{-1}$ (white-dashed path), 91 km s$^{-1}$ (blue-dashed path), 99 km s$^{-1}$ (gray-dashed path), 90 km s$^{-1}$ (orange-dashed path), 79 km s$^{-1}$ (purple-dashed path), and 86 km s$^{-1}$ (cyan-dashed path). Finally, we estimated the average (mean) speed and corresponding standard deviation which are 88.0 km s$^{-1}$ and 7.0 km s$^{-1}$. Finally, the mean projected jet upflow speed from AIA~304~{\AA} is 85$\pm$7.0 km s$^{-1}$. In Section~\ref{sect:evol_jet}, we showed that the jet is weakly visible in the 171~{\AA} filter and Fe~{\sc xviii} emission. However, the BNPC within the jet is clearly visible in all filters. Therefore, for the warm and hot temperature filters, we have selected a different path that goes through the BNPC (green dashed line in Figure \ref{fig:ht}(c) and (e)) to produce TD plots. Finally, the TD plots of the 171~{\AA} and Fe~{\sc xviii emission} are shown in Figure \ref{fig:ht}(d) and (f). Similar to the TD plot from AIA~304~{\AA}, the TD plot of AIA~171~{\AA} also shows multiple streaks, and again, we have drawn white-, orange-, pink-, and blue-dashed paths on AIA~171~{\AA} TD plot. The projected speeds are 101 km s$^{-1}$ (white-dashed path), 112 km s$^{-1}$ (orange-dashed path), 110 km s$^{-1}$ (pink-dashed path), and 119 km s$^{-1}$ (blue-dashed path). The mean speed is 110 km s$^{-1}$ while the standard deviation is 7.4 km s$^{-1}$. Hence, Based on the 171~{\AA} TD plot, the mean projected BNPC upflow speed is 110$\pm$7.4 km s$^{-1}$.\\

Multiple bright streaks in the TD plots (Figure \ref{fig:ht}(b), (d), and (f)) are indicated by white arrows in panel (b). Such multiple bright streaks, occurring during the lifetime of this jet, are clear indications of repeated plasma injections. These multiple streaks occur at regular intervals, implying that energy is released periodically within the event. This aspect is discussed in Section~\ref{section:jet_qpp} with the help of wavelet analysis.

\subsection{Jet thermal structure }
\label{section:thermal_nature}
\begin{figure*}
    \includegraphics[trim = 2.5cm 1.0cm 6.8cm 1.0cm,scale=1.0]{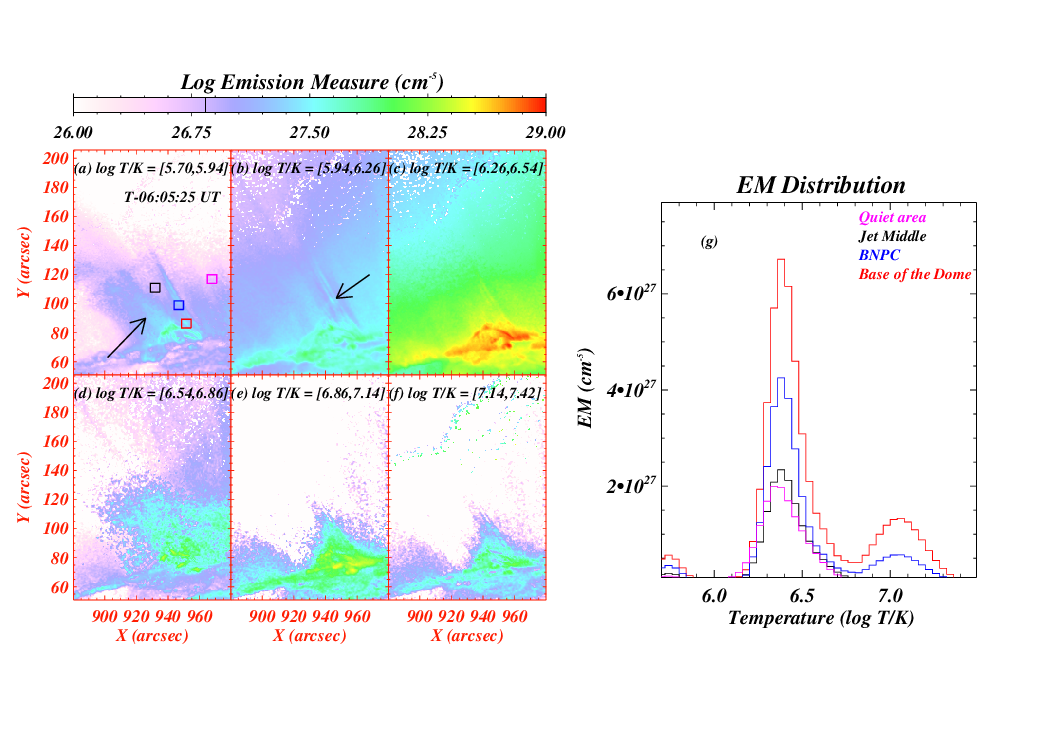}
    \caption{Left panels (a-f). The EM distribution of the jet at t = 06:05:25~UT (jet main phase) in six different temperature bins. The black arrows (shown in panels (a) and (b)) point to features described in the text. The right panel (g) shows the EM histogram (distribution) from four different regions outlined by boxes in panel (a). The histograms from these boxes are also labeled in panel (g)}.
    \label{fig:dem_jet_maps}
\end{figure*}

The full extent of the jet is mainly visible in cool filters (Figure~\ref{fig:jet_evol_cool}), but only weakly visible in the AIA~171~{\AA}, 131~{\AA}, and 94~{\AA} filters (Figures~\ref{fig:jet_evol_171} and~\ref{fig:jet_evol_hot}). To delve further into the multithermal nature of this event, we performed an emission measure analysis utilizing SDO/AIA images and software provided by \cite{Cheung2015}. The emission measure maps in various temperature ranges during the well-developed phase of this event (at t = 06:05:25~UT) are shown in Figure~\ref{fig:dem_jet_maps}.   
The lowest temperature bin, log T/K = 5.70-5.94 (Figure~\ref{fig:dem_jet_maps}(a)), shows emissions mainly in the BNPC within the jet, with a gap around the black box. We also see a faint jet above the cusp. The main body of the jet is weakly visible in the warm/hot filters of SDO/AIA (see Figures~\ref{fig:jet_evol_171} and~\ref{fig:jet_evol_hot}), but it is more prominent in the lower temperatures covered only by IRIS (Figure \ref{fig:jet_evol_cool}). Therefore, the emission from the main body of the jet is hardly present in the AIA-based emission measure maps of Figure~\ref{fig:dem_jet_maps}. However, the lowest temperature bin (log T/K = 5.70-5.94) contains significant emissions to the left of the jet (indicated by the black arrow in Figure~\ref{fig:dem_jet_maps}(a)), mainly due to the presence of the warm loops visible in AIA~171~{\AA} (Figure~\ref{fig:jet_evol_171}(f)). The next higher temperature bin (log T/K = 5.94-6.26; panel (b)) also shows a narrow zone of intense emission mainly around the BNPC (black arrow in (b)) and the dome. The emission measures in the highest-temperature bins (Figure~\ref{fig:dem_jet_maps}(c)-(f)) peak mainly in the vicinity of the dome, plus a little emission near the base of the jet.

We have chosen four small regions (Figure~\ref{fig:dem_jet_maps}(a)) for detailed EM analysis: in the middle of the jet (black box), in the BNPC (blue box), at the base of the dome (red box), and in a quiet area (magenta box). The EM histograms averaged over all pixels in each selected box are plotted in Figure~\ref{fig:dem_jet_maps}(g). The EM histogram from the middle of the jet (black) matches the EM histogram of the quiet area (magenta). Both histograms peak around log T/K = 6.4 so there is no enhanced emission in this temperature range from the jet; it is just background emission. 
In contrast, the EMs of the spire (blue histogram in Figure~\ref{fig:dem_jet_maps}(g)) and base (red histogram) of the BNPC within the jet also peak around log T/K = 6.4, but their magnitude is more than 2-3 times stronger than that of the black/magenta histograms.
The EM histograms from the base and BNPC spire also exhibit a second emission peak around log T/K = 7.1. The secondary emission is stronger at the base than in the spire. Hence, the BNPC within the jet mainly emits around log T/K = 6.4 with some fainter emission around log T/K = 7.1. The strong high-temperature EM from the base of the dome (red histogram in Figure~\ref{fig:dem_jet_maps}(g))  indicates that the dome heats as the event progresses, supporting the conclusion from the Fe~{\sc xviii} emissions discussed in Section~\ref{sect:evol_jet}.

\subsection{Quasi-periodic energy release}
\label{section:jet_qpp}
Investigation of the light curves from AIA~304~{\AA} and 171~{\AA} reveals quasi-periodic intensity variations in this jet event.
\begin{figure*}
    \mbox{
    \includegraphics[trim = 2.0cm 2.0cm 4.5cm 1.0cm,scale=0.8]{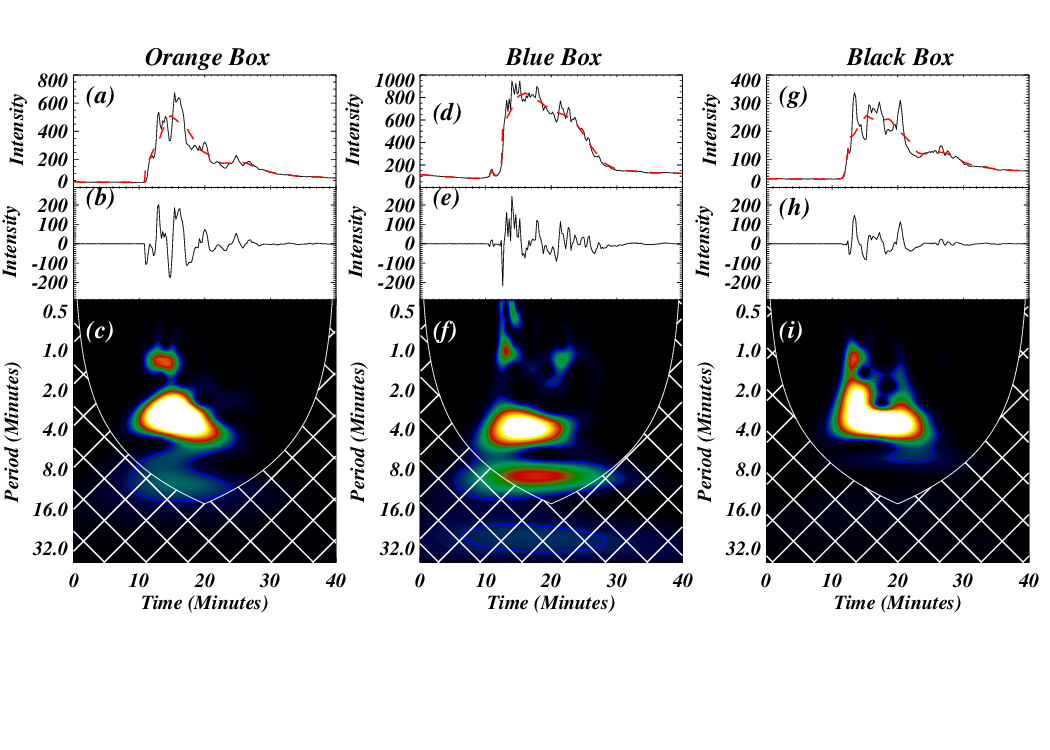}
    }
	\caption{Wavelet analysis of 304~{\AA} emission in three different locations (see boxes in Figure \ref{fig:jet_evol_171}(c)). The initial time t = 0  corresponds to t = 05:47:07~UT. The panel (a) shows the original light curve (black curve) and smoothed curve (red curve) from the orange box. While the panel (b) shows processed light curve (original curve-smoothed curve), and the corresponding wavelet power is shown in the panels. Similar analysis is shown for the blue (panels d, e, and f) and black boxes (panels g, h, and i).}
 \label{fig:qpp_304}
\end{figure*}  
We have selected three regions in Figure~\ref{fig:jet_evol_171}(c) to study the variability in this event: the spire of the jet (black rectangular box), the top of the dome where the jet is initiated (orange rectangular box), and the interaction region between the bright arcade and the right side of the dome (blue rectangular box). We have averaged the intensity over all pixels in each box to get three light curves from 304~{\AA} and three light curves from 171~{\AA}.

Before applying wavelet analysis to determine whether the episodic brightness variations were periodic, we processed the light curves as follows. Figure~\ref{fig:qpp_304} shows the original and pre-processed light curves along with the corresponding wavelet power maps. Panel (a) shows the averaged original light curve (black line) from the orange box along with the overplotted smooth curve (red dashed line), which was produced using a smoothing function with a window of 20 steps (240 seconds). The difference between the original light curve and the smoothed light curve is plotted in Figure~\ref{fig:qpp_304}(b). The results of applying wavelet analysis to this original-smoothed light curve are shown in panel (c). The white hatched area represents the cone of influence in the wavelet power map, while the solid blue line outlines the 95\% confidence level. Within the time range of 8-25 min, the most significant period is around 4 min. In addition to the major peak, there is a significant signal around 9 min, which mostly lies within the cone of influence.

Similarly, Figure~\ref{fig:qpp_304}(d) shows the original averaged light curve (black line) from the blue box overplotted by the smoothed curve (red dashed line), and panel (e) shows the difference between the two curves. As for the emission in the orange box, the corresponding wavelet power map in panel (f) displays significant power around 4 min and a longer patch of significant power around 9 min, all of which lie within the cone of influence. We have followed the same procedure for the light curve from the black box in the jet spire (panels (g)-(i)). The wavelet power map shows significant power around 4 min, as found for the other two boxes, but does not exhibit a longer period peak. 
\begin{figure*}
    \mbox{
    \includegraphics[trim = 2.0cm 2.0cm 4.5cm 1.0cm,scale=0.8]{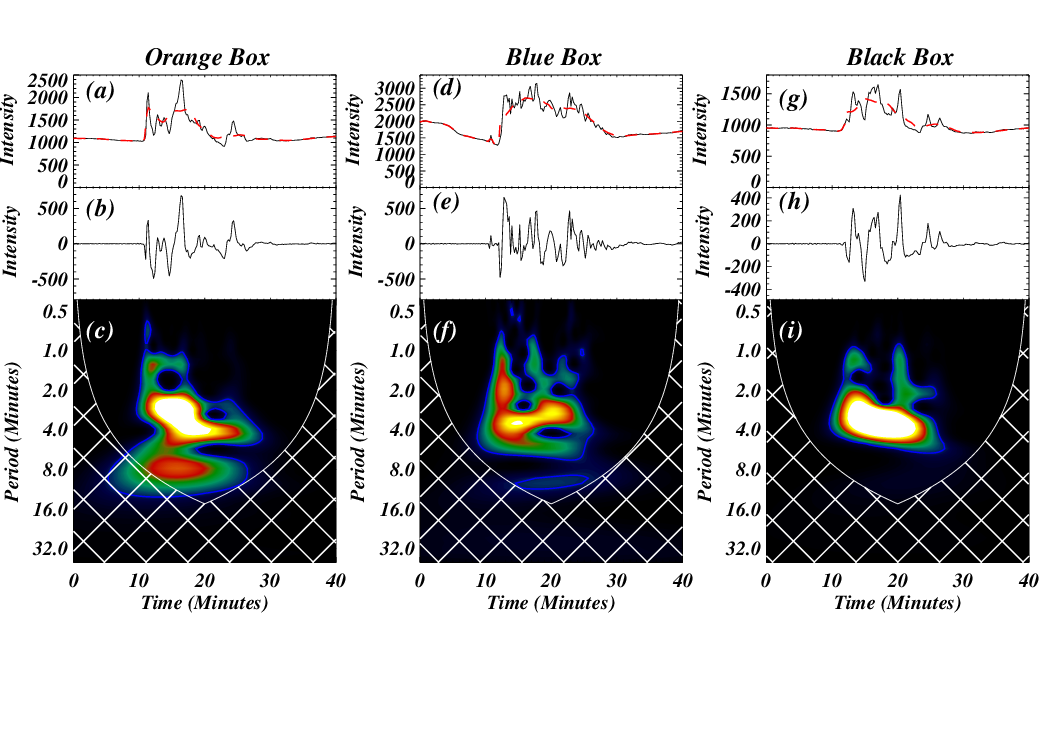}
    }
	\caption{Wavelet analysis of AIA~171~{\AA} intensity at three different locations (see boxes in Figure \ref{fig:jet_evol_171}(c)). The initial time t = 0 is the same as in Figure~\ref{fig:qpp_304}. Other details are same as described in Figure~\ref{fig:qpp_304}.} 
 \label{fig:qpp_171}
\end{figure*}  
We have investigated whether these quasi-periodic brightness variations also appear at higher temperatures by applying the same analysis to the same three boxes in the corresponding AIA~171~{\AA} image sequence. The original and pre-processed light curves, along with the wavelet power maps, are shown in Figure~\ref{fig:qpp_171}. The wavelet power map from the orange box (panel (c)) shows significant power around 4 min and 9 min, as in the 304~{\AA} channel. The blue and black boxes are dominated by a 4 min period (panels (f) and (i)).

As stated in Section~\ref{sect:obs_data}, IRIS observed the jet in 8-step raster mode. Because the jet was located at the left-bottom corner of the FOV, some parts of the jet were missing as IRIS completed 8 steps. Therefore, some areas within the black and orange boxes (Figure \ref{fig:jet_evol_171}(c)) in two alternative images (i.e., image steps 7 and 8) were not observed by IRIS. If the intensity time series were derived from these boxes, then the intensity time series would have missing points. Consequently we did not perform a wavelet analysis for the black and orange boxes.
\begin{figure*}
    \mbox{
    \includegraphics[trim = 2.0cm 2.0cm 4.5cm 2.0cm,scale=0.9]{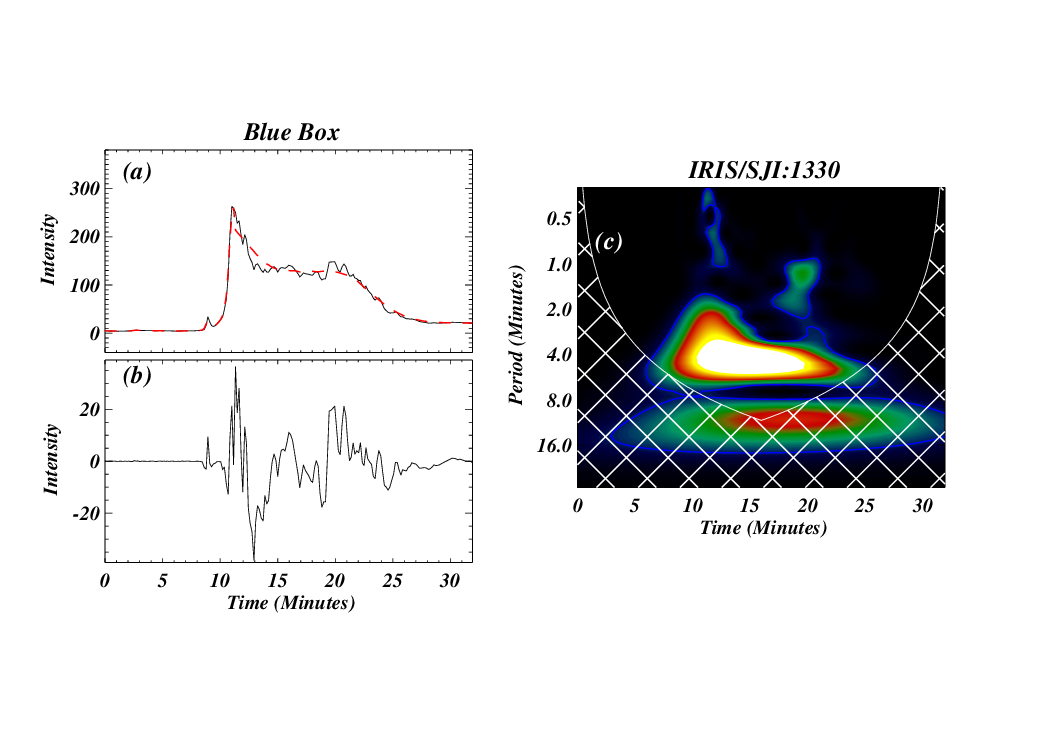}
    }
	\caption{Wavelet analysis of IRIS/SJI~1330~{\AA} intensity from only one location (see blue box in Figure \ref{fig:jet_evol_171}(c))}.
 \label{fig:qpp_iris}
\end{figure*}  

Because the blue box is located on the right edge of the dome, the jet was completely observed by IRIS in this area. Hence, we applied the wavelet analysis on the IRIS/SJI~1330~{\AA} light curve (Figure~\ref{fig:qpp_iris}) deduced from the blue box. Similar to Figures~\ref{fig:qpp_304} and~\ref{fig:qpp_171}, we have displayed the intensity time series (black curve in panel (a)) with its smoothed version (overplotted red dashed line in panel (a)), the detrended time series (panel (b)), and finally, the wavelet power map (panel (c)) in Figure~\ref{fig:qpp_iris}. Similar to the previous findings, the IRIS/SJI~1330~{\AA} intensity time series also exhibits quasi-periodic pulsations with a period of 4 minutes.
 
\section{Discussion and Conclusions} 
\label{sect:discussion}   
In this work, we have investigated the formation of a jet at the west solar limb using high-resolution imaging observations provided by IRIS and SDO/AIA. The underlying fan-spine magnetic topology is reflected in the AIA~171~{\AA} images (Figure~\ref{fig:jet_evol_171}(a)), and confirmed by a potential field extrapolation (Figure~\ref{fig:ref_fig}(b)). Multiple jets originated in the periphery of this active region, including several from the same bright point as the jet studied here, consistent with earlier observations of recurrent jets from bright points \citep{dhara2017, Kumar2019b, kumar2021}. 3D MHD simulations of coronal jets indicate that the basic fan-spine topology in this source region is unlikely to be significantly altered by recurrent jetting \citep[\eg ][]{pariat2010,Wyper2018,wyper2019}. Some closed flux will be opened and some open flux will be closed, depending on which part of the filament channel erupted. Flux conservation dictates that the total magnetic flux under the separatrix (dome) remains constant, so the separatrix can move around and distort but the total area under it won’t change. The system should relax toward its initial state between successive events, but some shear always remains inside the fan as long as the flare reconnection site is above the surface. How much shear or twist remains after each eruption, and how much is replenished between eruptions, are interesting questions, but we can't answer them for a limb event. In this case, the erupting western part was almost hidden behind the limb, making it impossible to track the evolution of the filament channel between two jet events. Ultimately the whole configuration shrank as the system relaxed after releasing energy, mass, and flux in the jet.

Several observational investigations have found that mini-filaments and mini-flare arcades exist in most coronal jet sources and are expelled with the jet  \citep{Innes2009, Sterling2015, Panesar2016, Kumar2018, Kumar2019a, Kumar2019b}, implying that mini-filament channels provide free energy for these eruptions in agreement with numerical studies of the breakout jet model \citep{Wyper2017, Wyper2018}. In the present event, a mini-filament was visible within the dome $\approx$40 min prior to the jet's initiation. The presence of the fan-spine topology and the mini-filament shows that the magnetic structure was conducive to eruptions, and that free energy was available in the form of a filament channel \citep[\eg][]{Wyper2017, Wyper2018, Kumar2019a}.

As per the magnetic breakout model, before the jet onset a current sheet formed around the null point in response to the rising stressed flux surrounding the mini-filament, creating a favorable configuration for breakout reconnection. In the analyzed event, brightenings appeared along the expanding dome and at its footpoints before the rising core containing the mini-filament reached the top of the dome (Figures~\ref{fig:jet_evol_cool}(b) and ~\ref{fig:jet_evol_171}(b)), which signify the beginning of weak breakout reconnection between the closed internal field and the open external field. As a result, the fan-spine structure became more clearly visible as it expanded (Figures~\ref{fig:jet_evol_cool},~\ref{fig:jet_evol_171}, and~\ref{fig:jet_evol_hot}(b)), in response to the extra magnetic pressure exerted by the stressed core flux. 
The weak reconnection in this breakout current sheet is accompanied by heating of the dome (described in Sections~\ref{sect:evol_jet} and \ref{section:thermal_nature}) and faint jetting (see Section~\ref{sect:evol_jet}).

The onset of weak breakout reconnection precedes the formation of the flare arcade beneath and towards the right side of the dome, as indicated in panels (b) of Figures~\ref{fig:jet_evol_cool},~\ref{fig:jet_evol_171}, and~\ref{fig:jet_evol_hot}. As the filament-channel flux rises, the flux beneath it eventually comes together and forms another current sheet, denoted the flare current sheet. When reconnection starts in the flare current sheet, a flux rope forms around the filament, and a flare arcade is created beneath it \citep[\eg][]{Forbes1996, Benz2008, Wyper2017}. Breakout and flare reconnection speed up, causing the flux rope to rise more rapidly until the flux above it has fully reconnected through the breakout current sheet. Once the flux rope itself comes into contact with the breakout current sheet, it is destroyed by reconnection with the external flux, and the entrained plasma is ejected with more force than in the previous interchange-reconnection phase. The event under study exhibited the key manifestations of this phase, signaling that explosive reconnection had begun (Figure~\ref{fig:jet_evol_cool}(b)): brightening of the dome surface and footpoints, simultaneous energetic expulsion of the jet and mini-filament plasma, and the onset of bright flare-arcade emission (Figures~\ref{fig:jet_evol_cool}, \ref{fig:jet_evol_171}, and \ref{fig:jet_evol_hot}(c-e)).  

In earlier jet modeling studies \citep[\eg ][]{Karpen2017} we found that the dominant reconnection site moves over the separatrix with time, and that multiple reconnection sites can coexist. Additional evidence for multiple reconnection sites comes from the numerous bright stripes in the TD plots (Figure~\ref{fig:ht}(b), (d,) and (f)) and the EUV light curves in all SDO/AIA channels (Figures~\ref{fig:qpp_304},~\ref{fig:qpp_171}, and~\ref{fig:qpp_iris}). The recurrent ejections and intensity peaks exhibit periods of $\approx$4 min throughout the 20-min jet lifetime, comparable to the periods reported for coronal jets \citep{Morton2012, Kumar2016, Mishra2023}, jetlets at the base of plumes and associated plumelets  \citep{Uritsky2021, Kumar2022, Kumar2023}, and for the peak in the power spectrum of coronal Alfv\'enic fluctuations \citep{Morton2019}. The implication of a connection between these ubiquitous quasi-periodic variations and a range of energy-release events on the Sun is intriguing, but requires further exploration beyond the scope of this paper. Our simulations also show that plasmoids form in both the breakout and flare current sheets \citep[\eg][]{Wyper2016b}. Patchy, intermittent, and plasmoid-generating reconnection all yield episodic energy release, producing episodic heating, particle acceleration, and flows. 

The studied jet was helical, due to the untwisting of the reconnected field lines. The twist on the flux rope inside the dome was transferred onto the external open field, producing twisted reconnected field lines. This twist could be enhanced by the precession of the reconnection site around the separatrix, or by multiple reconnection sites popping up within the current sheet extended along the separatrix. The mean projected speed of the main jet is around 88$\pm$7 km s$^{-1}$, while the BNPC mean projected jet speed is 110$\pm$7.4 km s$^{-1}$. In the breakout and resistive-kink jet simulations, the dense jet travels at roughly the sound speed (i.e., $\approx$150 km s$^{-1}$ at 1.0 MK plasma temperature), but an Alfv\'enic front precedes it at much higher speeds \citep{Karpen2017, Wyper2017}. Our observations do not reveal any Alfv\'enic front ahead of jet, most likely because the jet front is too thin and its density increase over the background is too small to be visible in EUV images. Although, the projected speed of the jet (i.e., 110$\pm$7.4 km s$^{-1}$ in the hot filter) is close to the sound speed. Therefore, the observed jet speeds (projected) are consistent with the breakout jet model. Note that neither the main jet nor the BNPC is due to chromospheric evaporation, which would come from the fan footpoints. 

The EM analysis (Figure~\ref{fig:dem_jet_maps}) confirms that the jet shows up in only cool emissions from plasma below log T/K = 5.70, while the BNPC, a bright and narrow plasma column within the jet, contains warm and hot plasma. Hence, the BNPC plasma exhibits a broader range of temperatures and is faster than the rest of the jet, indicating that the rate of energy release was higher during the BNPC.
The BNPC formed a few minutes after the triggering of the jet (Figures~\ref{fig:jet_evol_cool}(f), \ref{fig:jet_evol_171}(f) and \ref{fig:jet_evol_hot}(f)), and emanated from a different location (the right side of the dome), suggesting that additional reconnection started in a different site after the mini-filament eruption.   In numerical simulations of breakout jets, the breakout reconnection site begins near the initial null but shifts to below the erupting flux, essentially switching places with the flare reconnection site as part of the systemic relaxation toward a lower-energy state \citep{Wyper2018}. Therefore, it is possible that the primary reconnection site moved from the dome apex to the side of the fan closest to the flare arcade in the intervening few minutes. Alternatively, the flare arcade itself could have expanded until it encountered the closest section of the separatrix, driving a second, more energetic reconnection episode at the side of the dome. The available data do not definitively rule out either hypothesis.

We conclude that this eruptive event is consistent with the predictions of the magnetic breakout model for jets. The whole jet event highlights the complexity and variability of reconnection-driven solar activity, even in small events with relatively simple magnetic topology.

\begin{acks}
We gratefully acknowledge the reviewer for their constructive comments that improved the manuscript. IRIS is a NASA Small Explorer mission developed and operated by LMSAL with mission operations executed at NASA Ames Research Center and major contributions to downlink communications funded by ESA and the Norwegian Space Center. SDO observations are courtesy of NASA's SDO and the AIA and HMI science teams. The authors thank P. Wyper for valuable discussions. This research was supported by NASA's Heliophysics Guest Investigator (\#80NSSC20K0265), supporting research (\#80NSSC24K0264), GSFC Internal Scientist Funding Model (H-ISFM) programs, and the NSF SHINE program (Award Number \#2229336). Wavelet software was provided by C. Torrence and G. Compo, and is available at \href{http://paos.colorado.edu/research/wavelets/}{http://paos.colorado.edu/research/wavelets/}. Magnetic-field extrapolation was visualized with VAPOR (\href{(www.vapor.ucar.edu}{www.vapor.ucar.edu}), a product of the Computational Information Systems Laboratory at the National Center for Atmospheric Research.
\end{acks}



\section*{Supplemental Material}
1. IRIS/SJI 1330 \AA~ animation (Figure 2) \\
2. AIA 304, 171, and 94 \AA~ animation (Figures 3, 4, 5). The separation between each minor tick on the X and Y axes is 5$\arcsec$. \\  
 
\bibliographystyle{spr-mp-sola}
\bibliography{jet_pk_jt_pk_r2} 

\end{article} 

\end{document}